\documentclass[letterpaper]{jpconf}
\usepackage{graphicx}
\begin{document}

\title{Heavy Flavor Production at the electron-proton collider HERA}
\author{Klaus Urban}
\address{University of Heidelberg, Germany}

\begin{abstract}
An overview over the recent heavy flavor results of the H1 and ZEUS
collaborations is presented. Various techniques to tag the heavy
quark, which allow to explore different phase space regions, are 
employed. Predictions 
of pertubative QCD are compared to the charm and beauty production data. 
Charm and beauty fractions of the proton structure function $F_2$ are extracted.   
\end{abstract}

\section{Introduction}
The investigation of heavy quark production is one of the
main research topics at the ep collider HERA. 
Heavy quarks are predominantly produced via boson gluon fusion.
The mass of the heavy quark provides a large scale which allows a 
calculation of the parton scattering cross section 
$(g\gamma\rightarrow c\bar{c})$ in pertubative QCD.
The process $ep\rightarrow c\bar{c}(b\bar{b})$ is calculated by a 
convolution of the parton density functions of the proton (non pertubative part), 
the parton scattering cross section (pertubative part) and the 
fragmentation functions (non pertubative part), which
describe the transition from the heavy quark to a meson.
The measurement of heavy quarks is of interest since it allows a test of
pertubative QCD. Heavy quark provide a sizeable
contribution to the proton structure.
Cross section measurements for heavy quark production and
the extraction of the charm and beauty contribution to the proton 
structure function $F_2$, will be presented and
compared with theoretical pQCD predictions.

\section{Fragmentation}
\begin{figure}[t]
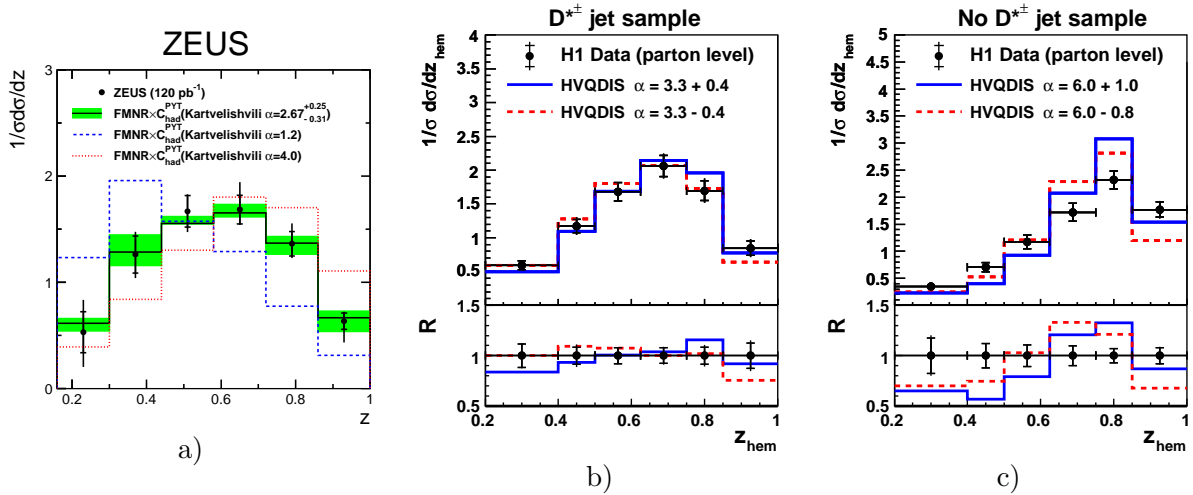

  \begin{minipage}[c]{.32\linewidth}
    \begin{center}
      \includegraphics[width=4.8cm]{DESY-08-209_4b.epsi} \\
      a)
    \end{center}
  \end{minipage}
  \begin{minipage}[c]{.32\linewidth}
    \begin{center}
      \includegraphics[width=4.8cm]{d08-080f7b.epsi}\\
      b)
    \end{center}
  \end{minipage}
  \begin{minipage}[c]{.32\linewidth}
    \begin{center}
      \includegraphics[width=4.8cm]{d08-080f10.epsi}\\     
      c)
    \end{center}
  \end{minipage}
  \hfill
  \begin{minipage}[c]{.99\linewidth}
    \begin{center}
      \caption{\it{
          \label{frag}
          {Measurement of the charm fragmentation function by the H1 and ZEUS collaborations 
            as a function of the variable $z$. Figure a) and b) shows the 
            measurement for the case that the $D^{\star}$ meson is measured together 
            with a jet. Figure c) shows the measurement for the case that no jet 
            is reconstructed in the event. The measurements are compared to 
            predictions provided by the NLO program HVQDIS.
          }}}
    \end{center}
  \end{minipage}
\end{figure}

The fragmentation function of charm quarks to $D^{\star}$ mesons
are determined as a function of $z$, which is defined as the ratio of the energy of the 
meson to the energy of the heavy quark. The fragmentation has 
been measured by the H1 and ZEUS collaborations \cite{fragH,Chekanov:2005mm}  where
the charm quark momentum is approximated by using two different methods. In one method 
the charm quark is approximated by the jet containing the $D^{\star}$-meson, this method 
works far above the production threshold. Within the other method the charm quark's energy
is approximated by the energy in the $D^{\star}$ meson's hemisphere; this method is applicable
close to the production threshold.
Figure \ref{frag} a) and b) shows the ZEUS and H1 fragmentation measurements as function a
of the variable $z$, for the case that the $D^{\star}$-meson is measured together with a jet.
These distributions are reasonably well described by the QCD model and similar
Kartvelishvili fragmentation parameters $\alpha$ \cite{Kartvelishvili:1977pi} are determind by H1 and ZEUS.
Figure \ref{frag} c) shows the H1 measurement, by using the hemisphere method, for the sample 
where no jet is found. In this sample a harder fragmentation parameter is needed to 
describe the data. No single parameter set describes the fragmentation in the whole
phase space.

\section{$D^{\star}$ meson production}
\begin{figure}[t]
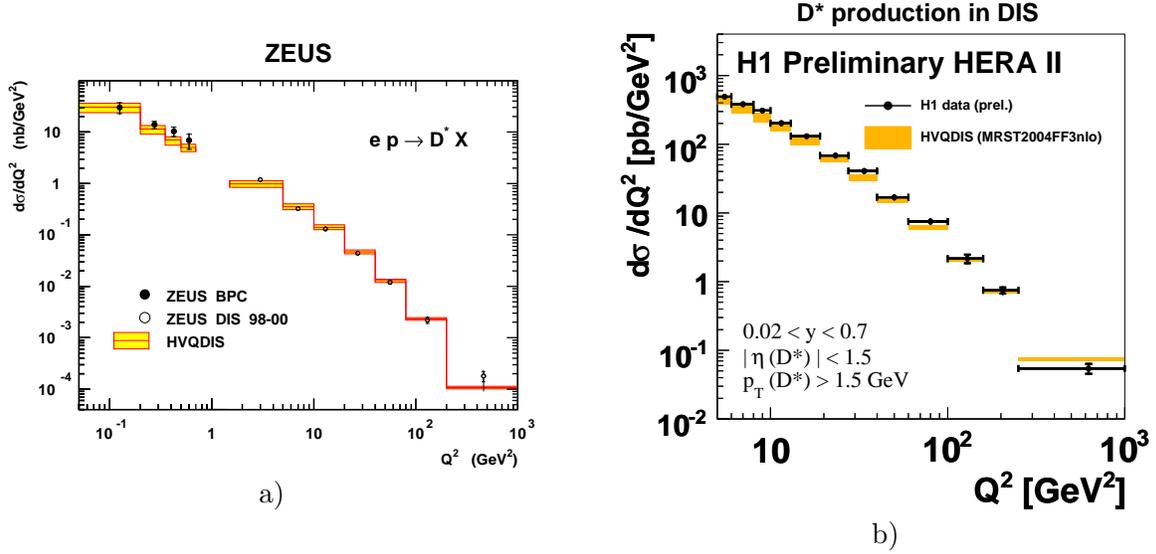

  \begin{minipage}[c]{.49\linewidth}
    \begin{center}
      \includegraphics[width=7.cm]{DESY-07-012_3.epsi}\\
      a)
    \end{center}
  \end{minipage}
  \begin{minipage}[c]{.49\linewidth}
    \begin{center}
      \includegraphics[width=7.cm]{H1prelim-08-074.fig8.epsi}\\
      b)
    \end{center}
  \end{minipage}
  \hfill
  \begin{minipage}[c]{.99\linewidth}
    \begin{center}
      \caption{\it{
          \label{dstarq2}
          {Measurement of the $D^{\star}$ cross section as a function of $Q^2$ performed by the H1
            and ZEUS collaborations.
            The measurements are compared to predictions provided by the NLO program 
            HVQDIS \cite{Harris:1997zq}.
          }}}
    \end{center}
  \end{minipage}
\end{figure}

Charm production can be identified by the presence of $D^{\star}$ mesons,
which are reconstructed in the decay channel
$D^{\star \pm} \rightarrow D^{0} \pi_{slow}^{\pm} \rightarrow K^{\mp} \pi^{\pm} \pi_{slow}^{\pm}$.
%are analyzed. 
These measurements are performed in a restricted 
$\eta$\footnote{The pseudorapidity $\eta$ is defined via the relation $\eta=-\ln(\tan \theta/2)$, where
$\theta$ is the polar angle with respect to the proton direction.} 
and $p_t$ range, which corresponds to the acceptance of the used detector devices.
The cross section measurement of $D^{\star}$ mesons as a function
of squared momentum transfer from the electron to the proton $Q^2$, measured in the kinematic region 
of deep inelastic scattering (DIS) by the H1 and ZEUS collaboration 
\cite{dsDIS,dsDIS_ZEUS} is shown in figure \ref{dstarq2}. The data is
compared to pQCD calculations provided by the program HVQDIS \cite{Harris:1997zq}.
A good agreement with the theory prediction is found over four orders of magnitude in $Q^{2}$.  
A similar measurement has been performed in photoproduction by the H1 collaboration \cite{dsgammap}.
The precision of these measurements is much higher than the accuracy of the NLO calculations.
The theoretical uncertainties are estimated by scale variations and correspond to missing
higher order terms.

\section{Measurement from semileptonic decays and inclusive lifetime tag}
The measurement of beauty production is based on the detection of semileptonic decays,
lifetime information or a combination of both methods. These methods allow the
measurement of charm production at the same time.
In contrast to the $D^{\star}$ analyses, both methods are sensitive 
to lower $p_t$ of the heavy quark. 
The ZEUS collaboration has performed a measurement of beauty production in the DIS kinematic regime 
based on two electrons in the final state $b\rightarrow ee$ \cite{dimuonZEUS} and 
a measurement 
which is based on semileptonic decays of beauty $b\rightarrow \mu x$ \cite{semiZEUS}. 
The H1 collaboration performed a similar measurement, based on semileptonic decays, 
in the photoproduction regime \cite{semiH}. 
A measurement which employs the lifetime information of the event has been performed by the 
H1 collaboration \cite{lifetime}. In all these measurements a good agreement  of
the beauty production with pertubative QCD is found.

\section{Extraction of $F_2^{c\bar{c}}$ and $F_2^{b\bar{b}}$}
\begin{figure}[t]
  \begin{minipage}[c]{.49\linewidth}
    \begin{center}
      \includegraphics[width=6.cm]{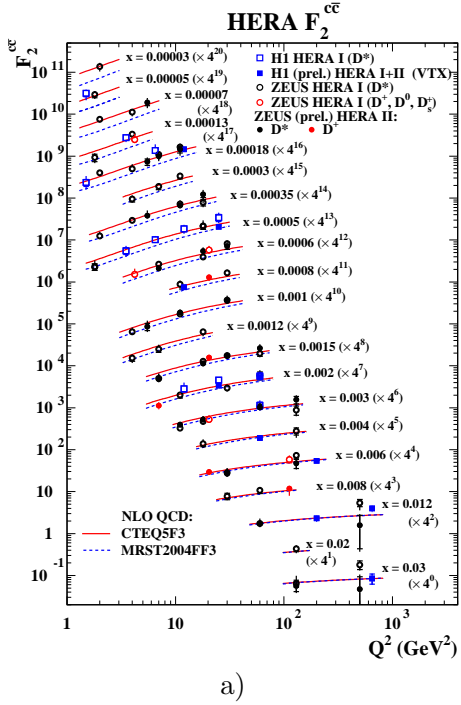}\\
      a)
    \end{center}
  \end{minipage}
  \begin{minipage}[c]{.49\linewidth}
    \begin{center}
      \includegraphics[width=5.cm]{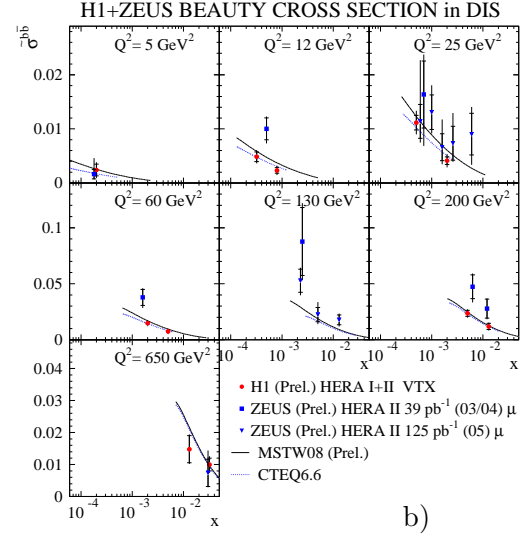}
      b)
    \end{center}
  \end{minipage}
  \hfill
  \begin{minipage}[c]{.99\linewidth}
    \begin{center}
      \caption{\it{
          \label{f2candf2b}
          {
            a) The structure function $F_2^{c\bar{c}}$ as a function of 
            $Q^2$ for various $x$ values. 
            b) The measured averaged reduced cross section
            $\tilde{\sigma}_{b\bar{b}}$ shown as a function of $x$ for different $Q^2$ values. The 
            inner error bars show the statistical error, the outer error bars represent
            the statistical and systematic errors added in quadrature. The data is compared to predictions 
            of pertubative QCD.
          }}}
    \end{center}
  \end{minipage}
\end{figure}

The structure functions $F_2^{c\bar{c}}$ and $F_2^{b\bar{b}}$ are defined in analogy
to $F_2$ and describe the fraction of charm (beauty) in the final state. In order to extract $F_2^{c\bar{c}}$ and $F_2^{b\bar{b}}$
from the measurements described above the cross sections are extrapolated to the  
full phase space. The extrapolation factors can be calculated
in pQCD, for which a good understanding of the fragmentation is required.
The structure functions are universal and hence can be used to compare experimental 
results from different analysis techniques.
Figure \ref{f2candf2b} shows a summary of the measurement of the proton structure function $F_2^{c\bar{c}}$ as
a function of $Q^{2}$ for different values of $x$ and $F_2^{b\bar{b}}$ in bins of $Q^2$ as a function of $x$. 
Clear scaling violations are observed towards large $Q^{2}$ at low values of $x$. 
A good agreement between the various analysis techniques and different data sets is found.
The data is compared to pQCD calculations, the precision of the charm data is high enough to 
resolve between different PDF sets at low $Q^{2}$ and $x$.
In general a good agreement with the pQCD prediction is found. 

\section{Acknowledgements}
I thank my H1 and ZEUS colleagues for their assistance in reporting these results.
I would also like to acknowledge and thank the organizers for their effort in arranging
this conference. I acknowledge financial support from the BMBF.


\newpage
\section*{References}

\begin{thebibliography}{9}
\bibitem{fragH} F.~D.~Aaron {\it et al.}  [H1 Collaboration], Eur.\ Phys.\ J.\  C {\bf 59}, 589 (2009) [arXiv:0808.1003].
\bibitem{Chekanov:2005mm}
  S.~Chekanov {\it et al.}  [ZEUS Collaboration],
  JHEP {\bf 0904}, 082 (2009)
  [arXiv:0901.1210 [hep-ex]].
\bibitem{Kartvelishvili:1977pi}
  V.~G.~Kartvelishvili, A.~K.~Likhoded and V.~A.~Petrov,
  Phys.\ Lett.\  B {\bf 78}, 615 (1978).
\bibitem{dsDIS} F.~D.~Aaron {\it et al.}  [H1 Collaboration], H1prelim-08-072 (2008),\\
{http://www-h1.desy.de/publications/H1preliminary.short\_list.html}.
\bibitem{dsDIS_ZEUS} ~S.~Chekanov {\it et al.} [ZEUS Collaboration],\\
  ZEUS Collaboration, ``Measurement of $D^*$ production in deep inelastic scattering and extraction of $F_2^{c \bar c}$'',
http://www-zeus.desy.de/physics/hfla/public/abstracts07/paper/f2charm\_zeus\_EPSPaper\_106.ps (2007).
\bibitem{Harris:1997zq} B.~W.~Harris and J.~Smith,
  Phys.\ Rev.\  D {\bf 57}, 2806 (1998) [arXiv:hep-ph/9706334].
\bibitem{dsgammap} F.~D.~Aaron {\it et al.}  [H1 Collaboration], H1prelim-08-073 (2008),\\
  {http://www-h1.desy.de/publications/H1preliminary.short\_list.html}.
\bibitem{dimuonZEUS} ~S.~Chekanov {\it et al.}  [ZEUS Collaboration],
  Phys.\ Rev.\  D {\bf 78}, 072001 (2008)
  [arXiv:0805.4390].
\bibitem{semiZEUS} ~S.~Chekanov {\it et al.} [ZEUS Collaboration], arXiv:0904.3487 (2009).
\bibitem{semiH} F.~D.~Aaron {\it et al.}  [H1 Collaboration], H1prelim-08-071, \\
  {http://www-h1.desy.de/publications/H1preliminary.short\_list.html}.
\bibitem{lifetime} F.~D.~Aaron {\it et al.}  [H1 Collaboration], H1prelim-08-173, \\
  {http://www-h1.desy.de/publications/H1preliminary.short\_list.html}.
\end{thebibliography}
\end{document}